\documentclass[amsmath,amssymb,aps,superscriptaddress,twocolumn,pra,longbibliography]{revtex4-1}
\usepackage{amsmath,amsfonts,amssymb,amsthm,graphics,graphicx}
\usepackage[colorlinks=true,citecolor=blue,linkcolor=blue,urlcolor=blue]{hyperref}
\usepackage{color}
\def\s{\sigma}
\def\om{\omega}

\def\da{\dagger}

\def\beq{\begin{equation}}
\def\eeq{\end{equation}}
\def\beqa{\begin{eqnarray}}
\def\eeqa{\end{eqnarray}}
\def\bal#1\eal{\begin{align}#1\end{align}}
\def\bfig{\begin{figure}}
\def\efig{\end{figure}}

\newcommand{\m}[1]{\mathcal{#1}}
\newcommand{\eq}[1]{Eq.~(#1)}

\newcommand{\fig}[1]{Fig.~#1}

\newcommand{\para}[1]{\left( #1 \right)}

\newcommand{\parc}[1]{\left\{ #1 \right\}}
\newcommand{\Phiad}{\left( \dot{\Phi}_i^\alpha \right)}
\newcommand{\Phiayd}{\dot{\Phi}_1^\alpha }

\begin{document}
\title{Resilience of the superradiant phase against $\mathbf {A^2}$ effects \\ in the quantum Rabi dimer}
\author{Yimin Wang}
\affiliation{Communications Engineering College, Army Engineering University, Nanjing 210007, China}
\author{Maoxin Liu}
\email{liumaoxin@bupt.edu.cn}
\affiliation{State Key Laboratory of Information Photonics and Optical Communications,
Beijing University of Posts and Telecommunications,
Beijing 100876, People's Republic of China}
\affiliation{School of Science, Beijing University of Posts and Telecommunications, Beijing 100876, People's Republic of China}
\affiliation{Beijing Computational Science Research Center, Beijing 100193, China}
\author {Wen-Long You }
\email{youwenlong@gmail.com}
\affiliation{College of Science, Nanjing University of Aeronautics and Astronautics, Nanjing, 211106, China}
\affiliation{School of Physical Science and Technology, Soochow University, Suzhou, Jiangsu 215006, China}

%
\author {Stefano Chesi}
\affiliation{Beijing Computational Science Research Center, Beijing 100193, China}

\author {Hong-Gang Luo}
\affiliation{School of Physical Science and Technology, Lanzhou University, Lanzhou 730000, China}
\affiliation{Beijing Computational Science Research Center, Beijing 100193, China}

\author {Hai-Qing Lin}
\affiliation{Beijing Computational Science Research Center, Beijing 100193, China}
\affiliation{Department of Physics, Beijing Normal University, Beijing, 100875, China}
\begin{abstract}
We explore the quantum criticality of a two-site model combining quantum Rabi models with hopping interaction.
Through a combination of analytical and numerical approaches, we find that the model allows the appearance of a superradiant quantum phase transition (QPT)  even in the presence of strong $\mathbf{A}^2$ terms, preventing single-site superradiance. In the two-site model the effect of $\mathbf{A}^2$ terms can be  surmounted by the photon delocalization from hopping, and a reversed superradiant QPT occurs as a consequence of the competition between $\mathbf{A}^2$ terms and the hopping interaction.  We characterize the phase diagram and scaling functions, and extract the critical exponents in the vicinity of the critical point,  thus establishing  the universal behavior of the second-order phase transition. Remarkably the effective hopping strength will be enhanced if more cavities are cascaded. We also prove that the multi-qubit counterpart of the quantum Rabi dimer, i.e., the Dicke dimer, has the same properties in beating the $\mathbf{A}^2$ effect. Our work provides a way to the study of phase transitions in presence of the $\mathbf{A}^2$ terms and offers the prospect of investigating quantum-criticality physics and quantum devices in many-body systems.
\end{abstract}
\date{\today}
\maketitle
\section{Introduction}
\label{sec_intro}

The superradiant phase transition (SPT) is one of the most fascinating
emergent phenomena in quantum optics
\cite{hepp1973aop,hioe1973pra,rzazewski1975prl,baumann2010n,bhaseen2012pra,baumann2011prl,
bastidas2012prl,keeling2010prl,torre2013pra,dallatorre2016pra,gelhausen2017pra,kirton2017prl,moodie2018pra,nagy2015prl,ritsch2013rmp,klinder2015prl,landig2016n}
.  Its history can be traced back to nearly half century ago,   when
the superradiant phase was originally predicted in the Dicke model (DM)~\cite{hepp1973aop,hioe1973pra}.
However, only two years later, Rza\.{z}ewski {\it et. al.} pointed out that such SPT may not exist in realistic cavity quantum electrodynamics (QED) systems due to the presence of a diamagnetic term proportional to the squared electromagnetic vector potential $\mathbf{A}^2$ ~\cite{rzazewski1975prl}.
 This `no-go theorem'  states that one cannot access the superradiant phase
at a finite temperature in  a  cavity QED system.
Furthermore, the no-go theorem also applies to the  zero-temperature  quantum phase transition (QPT) which might be induced by tuning the light-matter coupling~\cite{emary2003pre,hwang2015prl}.
 Although the SPT in cavity QED is out of reach, a remarkable experimental demonstration  of the superradiant QPT was exhibited in a cavity Bose-Einstein condensate
(BEC)
system~\cite{baumann2010n}, where the $\mathbf{A}^2$ term is negligible. This work triggered renewed interest in the studies on the SPT, both theoretically~\cite{bhaseen2012pra,baumann2011prl,bastidas2012prl,keeling2010prl,torre2013pra,dallatorre2016pra,gelhausen2017pra,kirton2017prl,moodie2018pra,nagy2015prl,ritsch2013rmp} and experimentally~\cite{klinder2015prl,landig2016n}.

 The occurrence of SPT has been especially debated in circuit QED, which is considered as one of the most promising platforms for demonstrating quantum optical phenomena~\cite{ridolfo_photon_2012,stassi2013prl,sanchez-burillo_scattering_2014,romero_ultrafast_2012,kyaw_scalable_2015,wang_holonomic_2016,liao2016pra,wang_ultrafast_2017,gu2017pr}. In particular,
 recent significant progress has allowed the  experimental realization of the so-called ultra-strong coupling (USC) regime and even the deep strong coupling (DSC) regime~\cite{niemczyk_circuit_2010,forn-diaz_observation_2010,chen_single-photon-driven_2017,yoshihara_superconducting_2017,forn-diaz_ultrastrong_2017,forn-diaz2019rmp,kockum2019nrp}.
 Due to the  advance of the fabrication and control capabilities,  the system parameters can be finely tuned within a much wider parameter range than traditional light-matter interacting systems~\cite{wang2018njp}.
 Therefore, such solid-state based quantum systems provide us the versatility to probe quantum criticality, e.g., the \emph{superradiant phase transition}. Whether the intrinsic limitations of cavity QED systems translate into analogous constraints for circuit QED systems is still in dispute~\cite{jaako2016pra,lu2018praa,lu2018prab}. It has been argued that the equivalent term $\mathbf{A}^2$  does not  prevent the occurrence of the QPT in certain circuit QED configurations~\cite{chen2007praa,lambert2009prb,nataf2010nc,nataf2010prl}. But those  arguments have faced disagreement  from both the general microscopic analysis~\cite{viehmann2011prl} and  specific examples~\cite{leib2014prl}. Moreover,
some theoretical proposals were put forward to experimentally settle the conflict~\cite{garcia-ripoll2015sr}.

 An interesting recent development on SPTs is the observation  that the  quantum Rabi model (QRM), with only a single atom, also exhibits superradiance in the limit where the frequency ratio of the qubit frequency $(\omega_q)$ to the resonator frequency $(\omega_r)$ tends to infinity~\cite{hwang2015prl,hwang2016prl,liu_universal_2017,larson2017jpamt,wang2018njp}, i.e., $\eta= \omega_q/\omega_r \rightarrow \infty$.  In this model,  a generalized form of universality has been established, which uncovers the equivalence between the many-body DM and the few-body QRM~\cite{liu_universal_2017}.  It was suggested that the relevant parameter for the SPT is $\eta N$, showing that the atom number $N$ in the DM has a direct correspondence with the frequency ratio $\eta$ in the QRM. Therefore, it is predictable that the QRM will also
 be subject to the no-go theorem~\cite{hwang2015prl}.

The aforementioned debate motivates us to investigate the SPT in  a broader class of Rabi models with $\mathbf{A}^2$ terms.  In this paper,  we propose a scheme to circumvent the $\mathbf{A}^2$ problem by involving the hopping interaction in a two-cavity quantum Rabi dimer (QRD).
We find  that the hopping between two cavities can compensate the $\mathbf{A}^2$ effect and  allows for a  superradiant phase. A modified version of the no-go theorem, including the hopping strength,  is presented. In contrast to the pristine QRM case, where the superradiant phase always appears by increasing the qubit-field coupling,   in certain parameter regimes a reversed scenario can occur, where the SPT takes place  by  decreasing the qubit-field coupling.
Moreover, the corresponding phase diagram and the scaling form in the vicinity of the quantum critical point are also investigated. To complete the discussion, the extended model with multiple atoms, i.e., the Dicke dimer, is also included. Our results suggest alternative approaches to realize the SPT in the presence of $\mathbf{A}^2$ terms. Furthermore, we expect our work  to stimulate experimental progress  in circuit QED systems.

Our paper is organized as follows. In Sec.~\ref{sec_model}, we introduce
the model and discuss the plausible parameter range prescribed by the stability of the two-cavity setup. In
Sec.~\ref{sec_qpd}, we analytically study the quantum phase transition of the two-cavity quantum Rabi dimer and portray the phase diagram. In Sec.~\ref{sec_scaling}, we
demonstrate the scaling and universality as well as the equivalence between Rabi dimer and Dicke dimer. In  Sec.~\ref{sec_dis} we discuss possible physical implementations  of the quantum Rabi dimer and finally  give a conclusion.

\section{Model and setup}
\label{sec_model}
\begin{figure}[tb!]
\begin{center}
\includegraphics[scale=0.33]{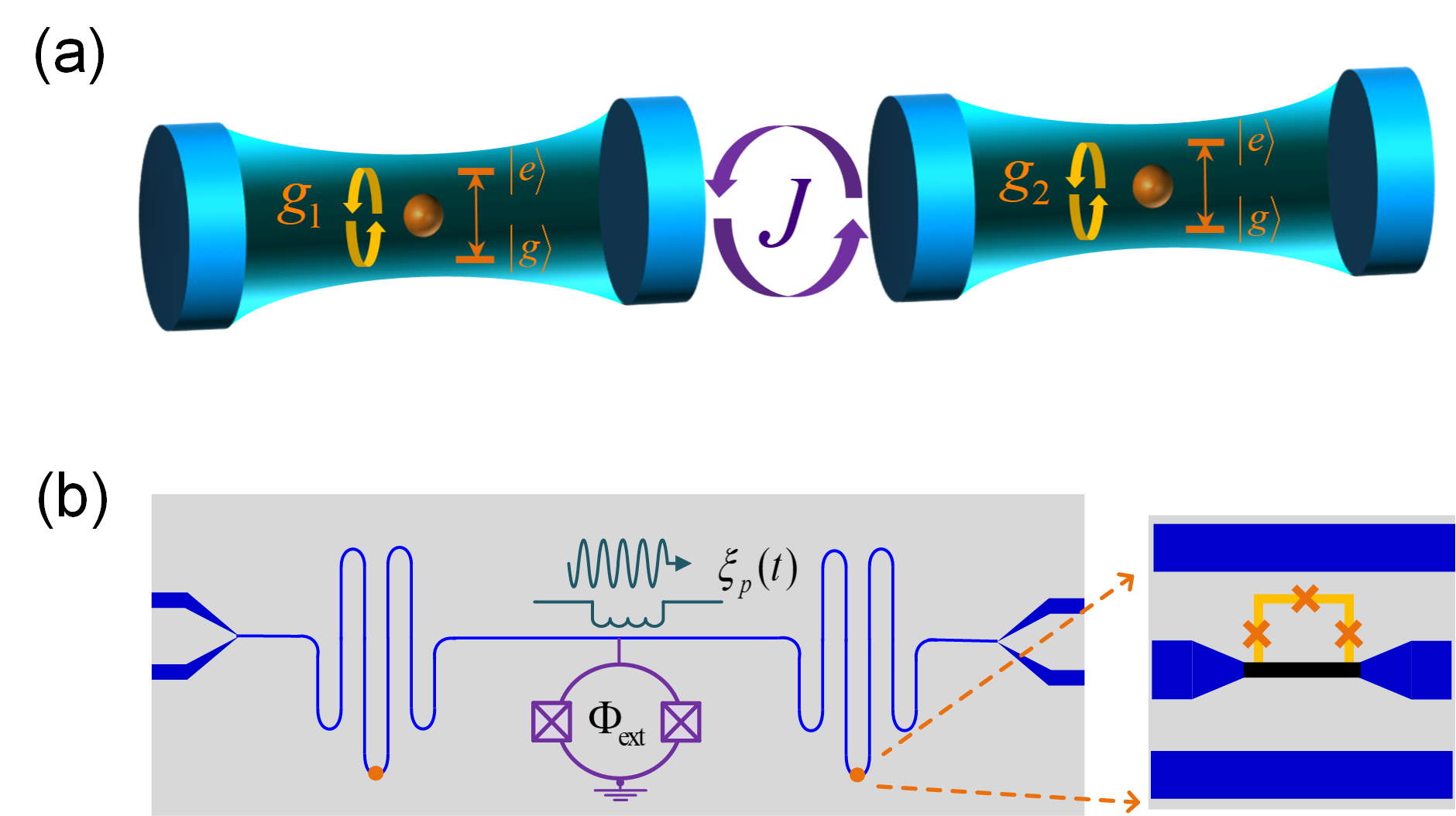}
\end{center}
\caption{(a) Schematic representation of a general quantum Rabi dimer, where the interaction between the two quantum Rabi systems is mediated through hopping coupling between cavities. (b) Circuit QED representation of the quantum Rabi dimer: two transmission line resonators are grounded through a SQUID. Each cavity is galvanically connected to a superconducting flux qubit (denoted by a orange dot) to achieve the ultrastrong (deep strong) coupling regime. The hopping interaction between the two resonators can be realized by modulating the external magnetic flux through the SQUID.}
\label{fig_setup}
\end{figure}

As is schematically depicted in \fig{\ref{fig_setup}}, the model we consider in this paper is the QRD with $\mathbf{A}^2$ terms. For simplicity and without loss of generality, we assume that  the   two cavities are identical.
The Hamiltonian for this system is described by
\bal
\label{eq_hd}
H =&\sum_{j = 1, 2} \left[\omega_r a_j^{\dagger}a_j+\frac{\omega_q}{2}\sigma_j^z + g \sigma_j^x(a_j + a_j^{\dagger}) + D(a_j + a_j^{\dagger})^2\right]\notag \\
&+J(a_1 + a_1^{\dagger})(a_2 + a_2^{\dagger}),
\eal
where $a_j$ ($a_j^\dag$) is the annihilation (creation) operator of the bosonic mode in $j$-th cavity with frequency $\omega_r$, $\omega_q$ is the qubit frequency, $\sigma_j^{z/x}$ are Pauli operators associated to a spin with spin-down state $|\downarrow_j \rangle$ and spin-up state $|\uparrow_j \rangle$, $g$ describes the coupling strength between the two-level system and the photonic field, $D$ denotes the amplitude of $\mathbf{A}^2$ term, and $J$ characterizes the strength of the intercavity hopping. The Hamiltonian in \eq{\ref{eq_hd}} respects a global $\mathbb{Z}_2$ symmetry of ${\m P} = e^{i \pi \hat{N}}$, where $\hat{N} = \hat{N}_1 + \hat{N}_2$ is the total excitation number and $\hat{N}_j = \s_{j}^+ \s_{j}^- + a_j^\da a_j$ is the number of excitations in the $j$-th QRM.

Considering the limit of negligible qubit-resonator coupling in the dimer Hamiltonian \eqref{eq_hd}, i.e., $g = 0$, the bosonic modes decouple from the qubits. The photonic excitations delocalize  due to  the hopping interaction. In this case, we obtain a quadratic Hamiltonian~\cite{jaako2016pra}
\bal
H_{\rm rdj}=& \omega_r  (a_1^\dagger a_1 + a_2^\dagger a_2) + J (a_1^\da + a_1 ) (a_2^\da + a_2 ) \notag\\
&+ D {(a_1^\da+a_1 )}^2 + D {(a_2^\da+a_2)}^2.
\label{eq_hrdj1}
\eal
This Hamiltonian can be diagonalized through a Bogoliubov transformation and rewritten as,
\bal
H_{\rm rdj} =\sum_{\alpha=\pm} \epsilon_\alpha c_\alpha^\dagger c_\alpha,
\label{eq_hrdj2}
\eal
where the two excitation frequencies are given by
\bal
\epsilon_\pm=\sqrt{\omega_r(\omega_r + 4 D \pm 2 J)}.
\label{eq_hrdjcon}
\eal
(More details can be found in Appendix A). It is  worth  noting that the value of $\epsilon_{\pm}$ should be real to ensure that the system is physically stable. This implies that the following condition needs to be satisfied,
\begin{equation}
\label{eq_stable}
\omega_r+4D\pm2J \geq 0,
\end{equation}
 which  sets the threshold for the stability of  the  cascaded cavities.

\section{Quantum phase transitions}
\label{sec_qpd}
To explore the ground-state properties of the model in \eq{\ref{eq_hd}}, we employ a Schrieffer-Wolff transformation $U=\exp[S]$ and obtain the effective Hamiltonian up to fourth order of $1/\sqrt{\eta}$ (see detailed discussions in Appendix B):
\bal
\label{eq_hd_sw}
H_{\rm eff}=& U^{\dagger}H U/\omega_q \notag \\
=&\sum_{j = 1,2} \left(\frac{a_j^{\dagger}a_j}{\eta}+\frac{1}{2}\sigma_j^z \right)+\frac{\tilde{J}}{2\eta}(a_1 + a_1^{\dagger})(a_2 + a_2^{\dagger})\notag \\
&+ \sum_{j = 1,2} \left[\frac{\tilde{g}^2(\sigma_j^z+\tilde{D})}{4\eta} (a_j + a_j^{\dagger})^2 - \frac{\tilde{g}^4\sigma_j^z}{16\eta^2} (a_j + a_j^{\dagger})^4\right],
\eal
where the generator is
\beq
\label{eq_s}
S=\sum_{j = 1,2}\left[-i\frac{\tilde{g}}{2\sqrt{\eta}}(a_j + a^\dag_j)\s_y + \frac{i}{6}{\left(\frac{\tilde{g}}{\sqrt{\eta}}\right)}^3 {(a_j + a^\dag_j)}^3\s_y\right].
\eeq
Here we have introduced a set of dimensionless parameters: $\tilde{g} = 2g/\sqrt{\omega_r \om_q}$, $\tilde{J}= 2J/\om_r$, and $\tilde{D} = D\om_q/g^2 $. At this point, the Hamiltonian \eqref{eq_hd_sw} can be reformulated in terms of renormalized coordinate and momentum operators, such that
\begin{eqnarray}
H_{\rm eff}&=&\sum_{j=1,2}\frac{1}{2}\Big\{\frac{p_j^2}{\eta^2 }+\sigma_j^z+\left[{\tilde{g}^2(\sigma_j^z+\tilde{D})+1}\right] x_j^2 - \frac{\tilde{g}^4\sigma_j^z}{2} x_j^4\Big\}
\nonumber \\
&+& \tilde{J} x_1 x_2,
\label{eq_hdxp}
\end{eqnarray}
where the quadratures are $x_j = (a_j^{\dagger}+a_j)/\sqrt{2\eta}$, and $p_j = i\sqrt{\eta} (a_j^{\dagger}-a_j)/\sqrt{2}$.

\emph{No-go theorem revisited.}
Having obtained the effective Hamiltonian \eqref{eq_hdxp}, we are able to describe
the SPT within the symmetry breaking scenario.
As a preliminary step, we revisit the no-go theorem for a single QRM. Assuming decoupled cavities,   i.e., $\tilde{J}=0$, \eq{\ref{eq_hdxp}} reduces to
\bal
H_{\rm eff}^{s} =\frac{\sigma^z}{2}+\frac{p^2}{2\eta^2}+ \frac{\tilde{g}^2(\sigma^z+\tilde{D})+1}{2}  x^2- \frac{\tilde{g}^4}{4}\sigma^z x^4.
\eal
Considering the limit $\eta \rightarrow \infty$, the kinetic term $p^2/(2\eta^2)$ can be neglected  and the low-energy effective potential  is obtained as
\begin{equation}
\label{eq_nogore}
E_0^s(x) =-\frac{1}{2}+ \lambda_s x^2 + \frac{\tilde{g}^4x^4}{4} +\cdots,
\end{equation}
where $\lambda_s$ $=$ ${[\tilde{g}^2(\tilde{D}-1)+1]}/2$.
The negative sign of the square-term coefficient ($\lambda_s$$<0$) indicates the occurrence of the superradiant phase.
However, as a consequence of  the  Thomas-Reiche-Kuhn (TRK) sum rule \cite{rzazewski1975prl}, the relation $D \geq g^2/\om_q$ always holds for a natural atom in the cavity. In this circumstance, the sign of $\lambda_s$ is always positive, which implies that the cavity QED system  cannot enter the superradiant phase  irrespectively  of the atom-resonator coupling strength. This is the well-known no-go theorem of the SPT in atomic systems.

In what follows, we will consider the effect of the intercavity hopping and discuss the QPTs in the Rabi dimer.
Similar to the QRM and again by taking the limit of $\eta \to \infty$ in \eq{\ref{eq_hdxp}}, the low-lying branch of the energy function for $\tilde{J}\neq0$ can be approximately expressed as
\begin{equation}
\label{eq_egs_x1x2}
E_0(x_1,x_2) =\sum_{j = 1, 2} \left(-\frac{1}{2} + \lambda_s x_j^2 + \frac{\tilde{g}^4}{4} x_j^4\right)
+\tilde{J} x_1 x_2+\cdots.
\end{equation}
One observes that the hopping interaction  hybridizes  the left mode ($x_1$) and the right mode ($x_2$). It is beneficial to rewrite \eq{\ref{eq_egs_x1x2}} in terms of the symmetric ($x_+$) and antisymmetric ($x_-$) normal modes with $x_{\pm} =(x_1 \pm x_2)/\sqrt{2}$,  thus  we have
\begin{equation}
\label{eq_egs_y1y2}
E_0(x_+,x_-)=-1+\lambda_{+} x_+^2+\lambda_{-} x_-^2+\frac{\tilde{g}^4}{8}\left(x_+^4 + 6x_+^2x_-^2+x_-^4\right),
\end{equation}
with coefficients
\begin{equation}
\label{eq_lambda}
\lambda_{{{\pm}}} = \lambda_s \pm \frac{\tilde{J}}{2} = \frac{[\tilde{g}^2(\tilde{D}-1)+1]}{2} \pm \frac{\tilde{J}}{2}.
\end{equation}
 From \eq{\ref{eq_egs_y1y2}}, we see that a SPT occurs when either one of the two coefficients $\lambda_{\pm}$ becomes negative. One finds that the inequality $\lambda_{\pm}  < 0$ might still have a solution even if $\tilde{D}>1$, which would prevent superradiance in a single cavity. In other words, it means that even for large $\mathbf{A}^2$-terms, there is still a possibility for the onset of the superradiant phase as long as the intercavity hopping amplitude $J$  is sufficiently strong. In contrast to a straightforward manner of overcoming the no-go theorem by reducing the strength of $A^2$-term,  here we present a novel approach to recover the superradiant phase  by introducing the hopping interaction between cavities.  Taking into account the stabilization constrain \eq{\ref{eq_stable}}, we find that the superradiant phase occurs in the following interval of $\tilde{J}$:
\begin{equation}
\label{eq_srcon1}
(\tilde{D}-1)\tilde{g}^2<|\tilde{J}|-1 \leq \tilde{D}\tilde{g}^2.
\end{equation}

\begin{figure}[tb!]
\begin{center}
\includegraphics[scale=0.35]{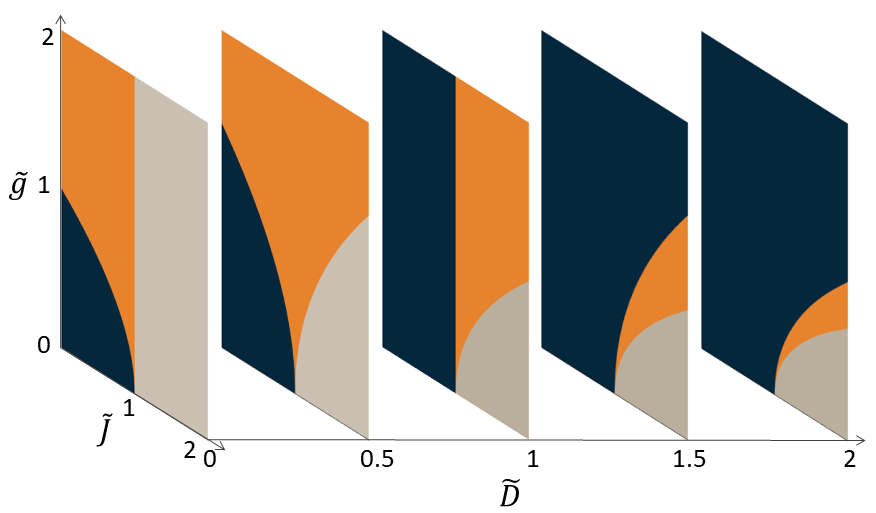}
\end{center}
\caption{Phase diagram in the ($\tilde{g}-\tilde{J}$) plane for different values of $\tilde{D}$. The black (orange) area stands for the normal (superradiant) phase, and the system is unstable within the gray region.}
\label{fig_pd1}
\end{figure}
\emph{Phase diagram.}  The critical point can be determined either by an analysis of the square-term coefficients, see \eq{\ref{eq_lambda}}, or by the derivative of the ground-state energy, obtained at $\eta \to \infty $ by minimizing \eq{\ref{eq_egs_y1y2}}.
The second-order derivative
shows a 
discontinuity at the  following  critical value of qubit-resonator coupling:
\beq
\tilde{g}_c = \sqrt{\frac{1-\tilde{J}}{1-\tilde{D}}}, 
\label{eq_gc}
\eeq
indicating the existence of a second-order QPT.

\begin{figure}[tb!]
\begin{center}
\includegraphics[scale=0.14]{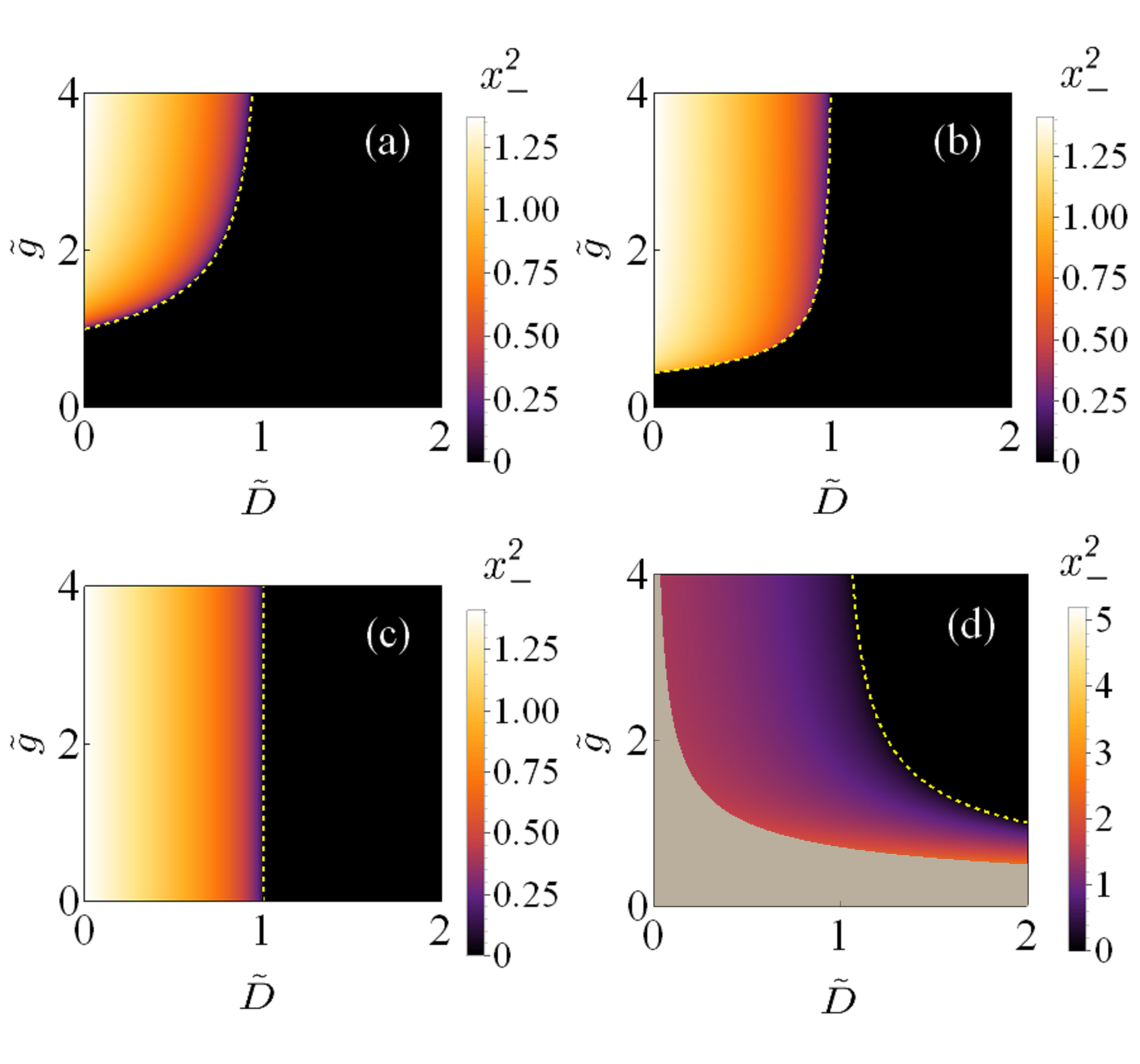}
\end{center}
\caption{The order parameter $\langle x_-^2 \rangle$ versus $\tilde{g}$ and $\tilde{D}$ for different values of hopping strength. (a) $\tilde{J} = 0$, (b) $\tilde{J} = 0.8$, (c) $\tilde{J} = 1$ and (d) $\tilde{J} = 1.5$. The dashed yellow lines depict the phase between the normal phase and the superradiant phase. The system becomes unstable within the gray region in (d).}
\label{fig_pd2}
\end{figure}

The evolution of this phase boundary with $\tilde{J}$ and $\tilde{D}$ is illustrated in Fig.~\ref{fig_pd1}. Furthermore, Fig.~\ref{fig_pd2} shows the expectation value $\langle x^2_- \rangle$in the ground state, which serves as order parameter.
In the superradiant phase with a negative $\lambda_-$, the expectation of $x_-^2$ can be obtained by $ \partial E_{0}\left(x_{+}=0, x_{-}\right)/\partial x_- =0$, leading to
\begin{equation}
\langle x_-^2 \rangle=\frac{2\sqrt{-\lambda_-}}{\tilde{g}^2}.
\end{equation}
Panel (a) of Fig.~\ref{fig_pd2} is in the absence of cavity hopping, i.e., $\tilde{J}=0$. One finds that there exists a QPT from the normal phase to the superradiant  phase by increasing $\tilde{g}$, and the critical value $\tilde{g}_c$ becomes larger upon increasing $\tilde{D}$ until a threshold value $\tilde{D}_c\equiv 1$.  For $\tilde{D} > \tilde{D}_c$, in agreement with the no-go theorem, the system can never reach the superradiant phase, even at very large atom-resonator coupling strength $\tilde{g}$.
When the intercavity hopping is weak, the  phase diagram  is qualitatively unchanged. As is shown in \fig{\ref{fig_pd2}}(b), the hopping strength plays a favorable  role in achieving the SPT, and  clearly  lowers the critical coupling strength $g_c$ for $\tilde{D} < \tilde{D}_c$. An extraordinary scenario arises when $\tilde{J}$ crosses  a threshold value $\tilde{J}_c \equiv 1$. As  shown  in \fig{\ref{fig_pd2}}(c), the system is frozen in the superradiant phase as long as $\tilde{D} < \tilde{D}_c$ and the ground state reverts to  the normal phase when $\tilde{D} > \tilde{D}_c$. Surprising and interesting features can also be  found  in the case when the hopping strength exceeds  $\tilde{J}_c$, when a reversed scenario is observed.
A typical example is displayed in \fig{\ref{fig_pd2}}(d) with $\tilde{J} = 1.5$.
In contrast to the  single-cavity  Rabi model, the superradiant phase is overwhelming and the normal phase becomes inaccessible for $\tilde{D} < \tilde{D}_c$. 
 For $\tilde{D} > \tilde{D}_c$, i.e., when the no-go theorem prevents the SPT in a single atomic cavity, the superradiant phase can still be realized and, remarkably, appears in the regime of weak light-matter coupling $\tilde{g}$. Conversely, the system enters the normal phase when $\tilde{g}$ is greater than the critical value $g_c$, exhibiting the opposite behavior of \fig{\ref{fig_pd2}}(a).
\section{Cascaded cavities}
After finishing the discussion of the two-cavity Rabi dimer, it is natural to bring multi-cavity case into scope. For cascaded cavities, the effective Hamiltonian reads
\begin{eqnarray}
\label{HeffLOBC}
H^{L}_{\rm eff}&=&\sum_{j=1}^{L}\frac{1}{2}\Big\{\frac{p_j^2}{\eta^2 }+\sigma_j^z+\left[{\tilde{g}^2(\sigma_j^z+\tilde{D})+1}\right] x_j^2 - \frac{\tilde{g}^4\sigma_j^z}{2} x_j^4\Big\}
\nonumber \\
&+& \tilde{J} \sum_{j=1}^{L-1}x_j x_{j+1}.
\end{eqnarray}
Similar to Eq. (\ref{eq_egs_x1x2}), an approximate low-lying branch of energy function can be obtained:
\begin{equation}
E_0^{L}(\{x_j\}) =\sum_{j = 1}^L \left(-\frac{1}{2} + \lambda_s x_j^2 + \frac{\tilde{g}^4}{4} x_j^4\right)
+\tilde{J} \sum_{j = 1}^{L-1}x_j x_{j+1}+\cdots.
\end{equation}
The square term can be concisely formulated in terms of matrix as
\begin{equation}
\label{eq_egs_yj}
E_0^{(2)}(\{x_j\})={\bf X}_L^TA_{L}{\bf X}_L,
\end{equation}
where ${\bf X}_L$ is a  column vector consisting of $L$ entries from $x_1$ to $x_L$, and
\begin{equation}
\label{eq_egs_yj}
{\bf A_L} =
\begin{pmatrix}
\lambda_s & \frac{\tilde{J}}{2} & & &\\
\frac{\tilde{J}}{2} & \lambda_s &\frac{\tilde{J}}{2} & &\\
 & \frac{\tilde{J}}{2} &\lambda_s &\ddots &\\
 & &\ddots &\ddots &\frac{\tilde{J}}{2}\\
 & & &\frac{\tilde{J}}{2} &\lambda_s\\
\end{pmatrix}.
\end{equation}
The normal modes ${\bf Y}_L$ is introduced by a unitary transformation ${\bf U}_L$ as
\begin{equation}
{\bf X}_L= {\bf U}_L{\bf Y}_L,
\end{equation}
satisfying
\begin{equation}
{\bf \Lambda}_{L} = {\bf U}_L^{\dagger}A_{L}{\bf U}_L.
\end{equation}
Thus, the square term of the energy function is diagonalized as
\begin{equation}
E_0^{(2)}(\{y_j\})={\bf Y}_L^T\Lambda_L {\bf Y}_L={\rm diag}\{ \lambda^L_1,\lambda^L_2,\cdots,\lambda^L_L\},
\end{equation}
where
\begin{equation}
\label{lambdaLj}
\lambda^L_j= \lambda_s+\left(2\cos k_j\right)\frac{\tilde{J}}{2},
\end{equation}
with the quasimomenta $k_j= j \pi/(L+1)$ ($j$=1, 2,$\cdots$,$L$) for the open boundary condition of the cascaded cavities. The minimum of $\lambda^L_j$ is
\begin{equation}
\label{minimumofLj}
\lambda^L_1= \lambda_s-\frac{\tilde{J}_{\rm eff}^{L}}{2},
\end{equation}
where $\tilde{J}_{\rm eff}^{L} = \left(2\cos\frac{\pi}{L+1}\right) \tilde{J}$.
A negative sign of $\lambda^L_1$ indicates an occurrence of the superradiant phase. Comparing to Eq. (\ref{eq_lambda}), the effective hopping $\tilde{J}_{\rm eff}^{L}$ here becomes larger than the hopping in the tow-cavity case as cavity number $L$ increases. We also note an interesting feature that the effective hopping (\ref{lambdaLj}) is unchanged when the periodic boundary condition is assigned for the cascaded cavities, i.e.,
\begin{eqnarray}
\label{HeffLPBC}
H^{P}_{\rm eff}&=&H^{L}_{\rm eff}
+ \tilde{J}  x_L x_{1}.
\end{eqnarray}
In this case, in Eq.(\ref{lambdaLj}) $k_j$=$-\pi$+$ 2 (j-1) \pi /L$ ($j$=1, 2,$\cdots$,$L$) for even $L$ and $k_j$= $-\pi$+ $ (2 j-1) \pi/L$ ($j$=1, 2,$\cdots$,$L$) for odd $L$. A slight odd-even effect can be observed for $L$ is not large. The enhancement of the effective hopping is beneficial to achieve the superradiant QPT by bridging more cavities but the underlying physics qualitatively remains unvaried.

%
%
\section{Finite-frequency scaling}
\label{sec_scaling}
As well as analyzing the QPT in the infinite limit of $\eta$, the finite-$\eta$ scaling provides further information of quantum criticality. Universal phenomena in the spin-boson interacting system have been widely explored by many researchers~\cite{hwang2015prl,liu_universal_2017}, such as critical exponents and the universality class.
An important unanswered question is whether the quantum Rabi dimer discussed here can exhibit synchronization and collective behavior involving complex interactions and mutual exchange.

\begin{figure}[tb!]
\begin{center}
\includegraphics[scale=0.5]{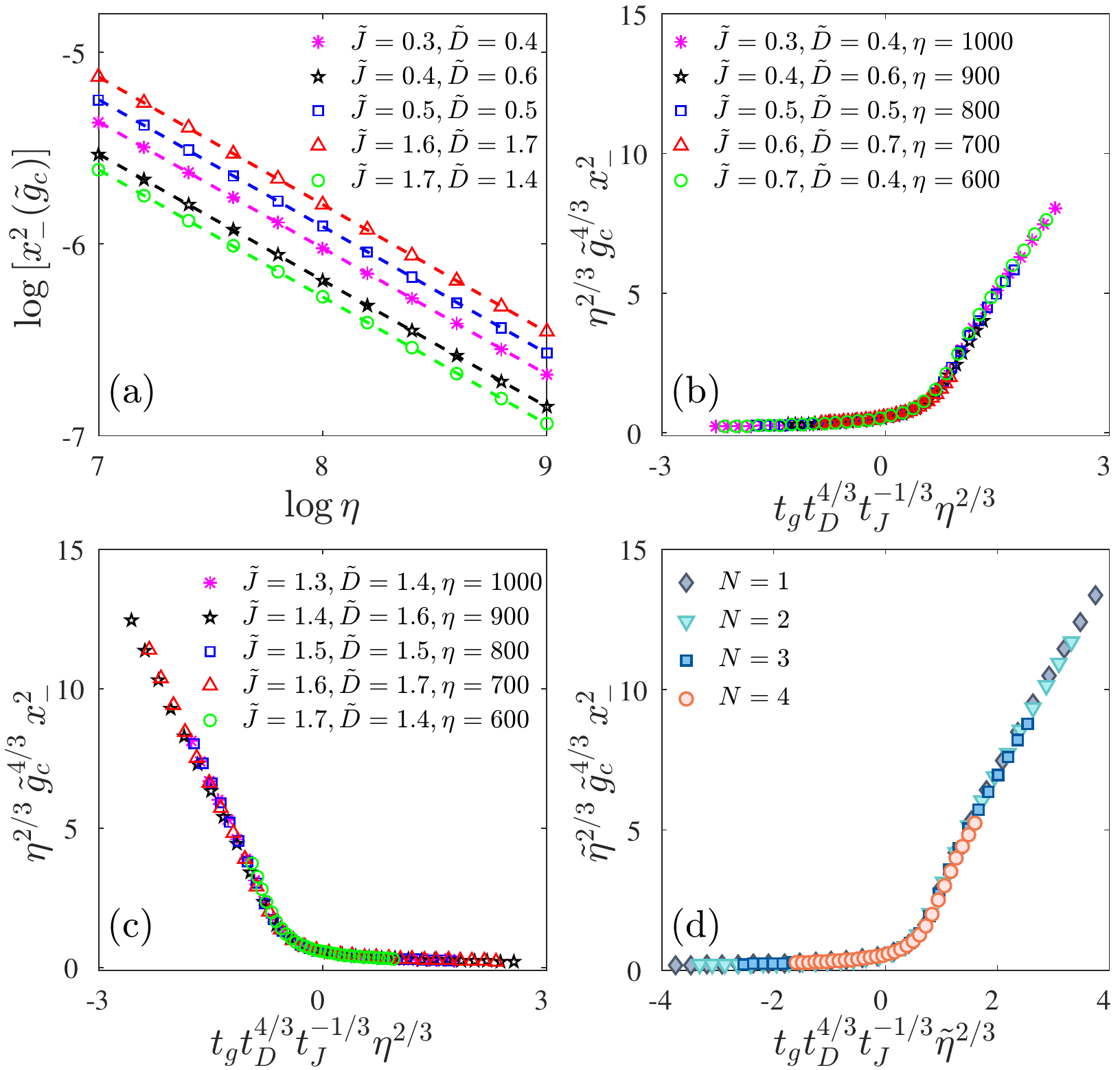}
\end{center}
\caption{(a) Log-log plot of the $x_-^2$ as a function of $\eta$ at the modified critical point $\tilde{g}_c$. With the linear fits of the numerical data, we find $x_-^2(\tilde{g}_c) \approx \eta^ {d_a^c}$ with $d_a^c \approx 0.33$ for different values of $\tilde{J}$ and $\tilde{D}$. The universal scaling of $x_-^2$ as a function of $t_g t_D^{4/3}t_J^{-1/3}\eta^{2/3}$ for different values of $\eta$, $\tilde{J}$ and $\tilde{D}$: (b) $\tilde{J}<1$ and $\tilde{D}<1$;  (c) $\tilde{J}>1$ and $\tilde{D}>1$. (d) Universal scaling function $x_-^2$ as a function of $t_g t_D^{4/3}t_J^{-1/3}\eta^{2/3}$ for different values of $N$ in the extended Dicke dimer model \eqref{eq_hdn}. Numerical data directly computed from \eq{\ref{eq_hdn}} collapses into a single line. The rest of the parameters are chosen as $\tilde{J} = 0.5$, $\tilde{D} =0.5$, $\eta = 2^9.$}
\label{fig_qrm2pXsJscaling}
\end{figure}

For this purpose, under the condition of finite $\eta$, the kinetic terms in \eq{\ref{eq_hdxp}} are no longer negligible, and then the effective low-energy Hamiltonian arrives at
\bal
H_{\eta}(x_+,x_-) =- \frac{1}{2\eta^2}\left(\frac{\partial^2}{\partial x_+^2}+\frac{\partial^2}{\partial x_-^2}\right)+E_0(x_+,x_-).
\label{eq_hscaling}
\eal
Note that since we are studying the scaling behavior around the critical point $x_{1}= -x_{2}$, or equivalently $x_{+} = 0$, $H_{\eta}(x_+,x_-)$ can be simplified to
\bal
H_{\eta}(0,x_-)=-1-\frac{1}{2\eta^2}\frac{\partial^2}{\partial x_-^2}+\lambda_- x_-^2 + \frac{1}{8}\tilde{g}^4 x_-^4,
\label{eq_hscaling2}
\eal
where the variable $x_{+}$ is removed.
In general the corresponding ground-state wave function $\phi_0$ depends on five variables of $\{ \omega_q,~\omega_r,~g,~D,~J\}$. However, by introducing appropriate scaling variables
\bal
u &= x_-\eta^{1/3}\tilde{g}_c^{2/3}, \\
v &= t_g t_D^{4/3}t_J^{-1/3}\eta^{2/3},
\eal
with $t_g= (\tilde{g}-\tilde{g}_c)/\tilde{g}_c$, $t_D= (\tilde{D}-\tilde{D}_c)/\tilde{D}_c$ and $t_J= (\tilde{J}-\tilde{J}_c)/\tilde{J}_c$ being the reduced couplings, the wave function $\phi_0$ can be described straightforwardly by the following equation in terms of two independent variables $u$ and $v$,
\begin{equation}
\label{eq_xpsa2_phigs3}
\left(-\frac{1}{2}\frac{\partial ^2}{\partial u^2}- vu^2+\frac{u^4}{8}\right)\phi_0(u,v)= \m{E}_0(v) \phi_0(u,v).
\end{equation}
One can readily notice that after considering multi-degree of freedom in the coupled cavities the universal eigenfunction \eqref{eq_xpsa2_phigs3} is intact comparing that of single cavity. The underlying mechanism originates from the fact that the added interactions still obey the $Z_2$ symmetry. To this end, the ground-state energy of \eq{\ref{eq_hscaling2}} is given by
\bal
E_G\simeq-1+\eta^{-4/3}\tilde{g}_c^{4/3} \m{E}_{0}(v )+\ldots.
\eal
Meanwhile, the universal scaling laws of different observables can be easily derived from the general form of the ground-state wave function $\phi_0(u,v)$. At this stage, we employ a general scaling ansatz for relevant observables, in the form of
\begin{align}
\label{eq_scaling_general}
Q=&\eta^{-\beta_Q /\nu} \tilde{g}_c^{\alpha_Q/\nu_{\kappa} }\tilde{Q}\left(t_g t_D^{4/3}t_J^{-1/3}\eta^{2/3}\right).
\end{align}
According to \eq{\ref{eq_xpsa2_phigs3}}, the universal scaling function of the fields position quadrature operator $X_{2,n}(v)$ can be obtained by taking the expectation values of $x_-^{2n}$ over the ground state $\phi_0(u ,v )$,
\begin{align}
\label{eq_scaling_x0rdiy2}
x_-^{2n}=&\eta^{-2n/3}\tilde{g}_c^{-4n/3}\,X_{2,n}\left(v\right).
\end{align}
To confirm the validity of the scaling function \eqref{eq_scaling_x0rdiy2}, we numerically calculate the finite-$\eta$ scaling around the critical point.
Taking the logarithm on both sides of \eq{\ref{eq_scaling_x0rdiy2}}, we get
\begin{align}
\label{eq_unlinear}
\ln x_-^{2n}=&-2n/3 \ln \eta -4n/3 \ln \tilde{g}_c + \ln X_{2,n}\left(v\right).
\end{align}
Pinning exactly at the critical point $t=0$ (thus $v = 0$), the scaling function associated with position quadrature $X_{2,n}\left(v\right)$ becomes a constant, that is,
\begin{align}
\label{eq_linear}
\ln x_-^{2n}|_{t=0}=&-2n/3 \ln \eta + \rm{const}.
\end{align}
The linear relation between $\ln x_-^{2}$ and $\ln \eta$ is elucidated in \fig{\ref{fig_qrm2pXsJscaling}}(a). According to \eq{\ref{eq_linear}}, the slope of the linear log-log relation equals to the critical exponent $1/\nu$, indicating that $1/\nu$ is $2/3$. In the vicinity of the critical point, the linearity of numerical values of $\ln x_-^{2n}$ and $\ln \eta$ breaks down. Note that in \eq{\ref{eq_scaling_general}} two types of new critical exponents $\nu_{\kappa}$ and $\alpha_{Q}$ are introduced. Figures \ref{fig_qrm2pXsJscaling}(b) and (c) show  that, if the value of critical exponents are properly chosen, e.g., $\nu=3/2$, $\nu_{\kappa} = 3$ and $\alpha_{Q} =1$, curves with different scales of $\{\eta$, $\tilde{D}$, $\tilde{J}\}$ collapse into a single curve which corresponds to the scaling function of $X_{2,n}(v)$ in \eq{\ref{eq_scaling_x0rdiy2}}.


\emph{Universality and equivalence.}
Closely related to the QRM is its $N$-particle generalization --- the Dicke model, which has long been known for observing SPTs in the
thermodynamic limit
$N \rightarrow \infty$
~\cite{baumann2010n,klinder2015prl,landig2016n}
. Ref.~\cite{liu_universal_2017} proves that the frequency ratio $\eta$ for the QRM
plays an equivalent role as the particle number $N$ in the Dicke model.
That is to say, in the general $N$-qubit model, an essential parameter characterizing the energy scale across the QPT is $\tilde{\eta}= N \eta$.
It is natural and necessary to investigate the critical phenomena in the Dicke dimer, whose Hamiltonian is described as
\begin{eqnarray}
\label{eq_hdn}
H_N &=&\sum_{j = 1, 2} \Big[\omega_r a_j^{\dagger}a_j+\frac{\omega_q}{2}\sum_{i = 1}^N \sigma_{i,j}^z + \frac{g}{\sqrt{N}} \sum_{i = 1}^N \sigma_{i,j}^x(a_j + a_j^{\dagger}) \notag \\
&+&D(a_j + a_j^{\dagger})^2\Big]+J(a_1 + a_1^{\dagger})(a_2 + a_2^{\dagger}).
\end{eqnarray}
Following the similar strategy, we compute the universal scaling function of the position quadrature $x_-^2$ as a function of $t_g t_D^{4/3}t_J^{-1/3}{\tilde{\eta}}^{2/3}$.
The numerical results for different values of $N$ are shown in \fig{\ref{fig_qrm2pXsJscaling}}(d). Despite the differences of qubit number $N$, the expectation values of the position quadrature also collapse onto a single curve when appropriately scaled. Therefore, the equivalence between the Rabi dimer and the Dicke dimer is established.

\section{Discussions and Conclusions}
\label{sec_dis}
In principle, a straightforward implementation of the proposed model could be realized in cavity QED systems, where the $\mathbf{A}^2$ term naturally appears. The required quadratic coupling between different optical cavity modes usually results from the overlap of their spatial distribution, which normally can be obtained with a partially reflecting mirror \cite{hartmann2006np}.
However, typical values of the qubit-resonator coupling and the hopping interaction in optical cavity QED systems are too weak to demonstrate the QPT. A reasonable and practical solution would be quantum simulation, where an effective QRM with extremely large coupling at each site can be realized~\cite{gutierrez-jauregui2018aq}.

An alternative and promising candidate for demonstrating our investigations is the circuit QED system.
As schematically depicted in \fig{\ref{fig_setup}}(b), the ultrastrong and the deep strong coupling regimes of each Rabi site may be achieved
by implementing a longer and thinner shared line between a flux qubit and a microwave resonator~\cite{niemczyk_circuit_2010,yoshihara_superconducting_2017}. A tunable hopping rate can be realized by the external flux threaded through the superconducting quantum interference device (SQUID). Interestingly, $\mathbf{A}^2$ terms of cavity in each site may come naturally with the hopping due to the parametric processes induced by the SQUID connecting the two cavities \cite{felicetti2014prl,wang_holonomic_2016}. Therefore, although there still exists a debate on the no-go theorem in circuit QED systems, this typical configuration may act as a good testbed to manifest our theoretical predications in surmounting the $\mathbf{A}^2$ problem in the phase transition of the QRD. Although with regular parameters in circuit QED, the hopping strength is around a few megahertz, a possibility of reaching up to and going beyond $\sim1.2$~GHz can be expected~\cite{peropadre2013prb}.
Moreover, a hopping coupling rate of $13\%$~($\sim700$~MHz) has been demonstrated experimentally by increasing the shared electrical length~\cite{haeberlein2013acp}, and further enhancement to boost towards the corresponding ultrastrong hopping can be forecasted with inhomogeneous resonators interrupted by Josephson junctions~\cite{peropadre2013prb}. Besides, the strengths of the hopping interaction and the $\mathbf{A}^2$ terms can be modified by introducing parametric drives onto the resonators~\cite{lu2015prla,qin2018prl,leroux2018prl}. These are believed to be essential to circumvent the no-go property of the $\mathbf{A}^2$ terms in the QRD.

In this work, we consider a quantum Rabi dimer consisting of two Rabi cavities with the $\mathbf{A}^2$ term and intercavity hopping. We utilize Landau theory to
investigate the quantum phase transition and the associated critical behavior. Within the spontaneous-symmetry-breaking scenario, the emergence of the superradiant phase transition can be intuitively identified by the sign change of the square-term coefficient of the effective bosonic field Hamiltonian. In a single-site Rabi model, a quantum phase transition from the normal phase to the superradiate phase takes place with increasing the atom-resonator coupling $\tilde{g}$, and the critical value $\tilde{g}_c$ swells upon until diverges as $\tilde{D}$ increases to $\tilde{D}_c$.
For $\tilde{D}$ $>\tilde{D}_c$, the routine to the superadiant phase is forbidden, as is the notorious no-go theorem. Turning on the intercavity hopping between two cavities, we find that the bosonic hopping effectively compensates the $\mathbf{A}^2$ effect. The intercavity hopping reduces the value of critical coupling $\tilde{g}_c$, so the superradiant phase can be realized by a smaller qubit-resonator coupling. Intriguingly, for $\tilde{J}>\tilde{J}_c$, the roles of both phases are interchanged. To be more precise, the superradiant phase is currently overwhelming with a weak qubit-resonator coupling, and even invincible for $\tilde{D} < \tilde{D}_c$. To achieve the superradiance, the hopping integral
can compensate for the light-matter coupling. Interestingly, if more cavities are involved, the underlying picture is almost unchanged, except that the effective hopping strength is enhanced. It opens an innovative door to experimentally realize the superradiant phase.

In the vicinity of the critical point, we extract a scaling form involving multi-parameter by both analytical and numerical calculations. Thus, the critical exponents and universality class have been determined. Although this analysis is explicitly illustrated in the two-cavity case, an equivalent scaling analysis can be readily obtained for an arbitrary number of the cascaded cavities if Eq. (\ref{minimumofLj}) is applied. Moreover, we also consider the Dicke dimer by increasing the atom numbers in each cavity. Through a general universal scaling form of various observables with more critical exponents, we unveil the equivalence of the quantum critical behaviors between the single-atom Rabi dimer and the multi-atom Rabi dimer after calibrating energy scale.
We expect that our work brings new insight of the critical behavior of the superradiant phase transition, in view of an impressive ongoing progress of technologies to design a minimum system for multi-cavity setup.

\section*{Acknowledgments}
\label{sec_ack}
The authors appreciate the valuable help from Zeng-Qiang Yu. This work was supported by the Natural Science Foundation of China (NSFC) (Grant Nos. 11604009, 11674139, 11404407, 11474211, 11834005) the Program for Changjiang Scholars
and Innovative Research Team in University, China (Grant No. IRT-16R35), and the Fundamental Research Funds for the Central Universities. S. Chesi acknowledges support from the National Key Research and Development Program of China (Grant No. 2016YFA0301200) and NSFC (Grants No. 11574025, No. 11750110428, and No. 1171101295). W.L is appreciative of support from the start-up fund of Nanjing University of Aeronautics and Astronautics. H. Q. Lin thanks support from NSFC 11734002. We also acknowledge financial support from NSAF U1930402 and computational resources from the Beijing Computational Science Research Center.



\renewcommand{\theequation}{A-\arabic{equation}}
\setcounter{equation}{0}  

\renewcommand{\thefigure}{A\arabic{figure}}
\setcounter{figure}{0}  

\appendix

\section*{Appendix A: Diagonalization through Bogoliubov transformations}
\label{sec_app1}
To diagonalize the quadratic Hamiltonian displayed in \eq{\ref{eq_hrdj1}} using Bogoliubov transformations, we define a new pair of quasi-particle operators
\bal
\gamma = \alpha_1 a_1 + \alpha_2 a_2 - \beta_1 a_1^\dagger - \beta_2 a_2^\dagger, \\
\gamma^\dagger = \alpha_1 a_1^\dagger + \alpha_2 a_2^\dagger - \beta_1 a_1 - \beta_2 a_2,
\label{eq_gamma}
\eal
where $\alpha_{1(2)}$ and $\beta_{1(2)}$ are complex numbers.

The Bogoliubov transformation is the canonical transformation mapping the operators $a_{1(2)}$ and $a_{1(2)}^\dagger$ to $\gamma$ and $\gamma^\dagger$. To guarantee that the transformation is canonical, the commutator is evaluated to has the bosonic commutation relation,
\bal
[\gamma, \gamma^\dagger ] = 1,
\eal
and it is then evident that the coefficients $\alpha_{1(2)}$ and $\beta_{1(2)}$ must satisfy the following rule
\bal
\alpha_1^2 + \alpha_2^2 - \beta_1^2-\beta_2^2=1.
\label{eq_cons}
\eal
The quadratic Hamiltonian in \eq{\ref{eq_hrdj1}} can then be diagonalized and written as
\bal
H_{\rm rdj} = \epsilon \, \gamma^\dagger \gamma + (\text{the part commutes with } \gamma \text{ and } \gamma^\dagger),
\eal
with respect to the new annihilation and creation operators.
The commutation relation for the bosonic operator $\gamma$ is
\bal
[\gamma, H_{\rm rdj}] = \epsilon \,\gamma.
\label{eq_eom1}
\eal
Substituting \eq{\ref{eq_hrdj1}} into \eq{\ref{eq_eom1}}, and pairing the coefficients in the left and right sides of \eq{\ref{eq_eom1}}, we arrive at a set of equations
\bal
(\omega_r + 2D) \alpha_1 + J \alpha_2 + 2D \beta_1 + J \beta_2 = \epsilon \alpha_1, \\
J \alpha_1 + (\omega_r + 2D) \alpha_2 + J \beta_1 +2D  \beta_2 = \epsilon \alpha_2 , \\
2D \alpha_1 + J \alpha_2 +  (\omega_r + 2D) \beta_1 + J \beta_2 = -\epsilon \beta_1, \\
J \alpha_1 + 2D \alpha_2 + J \beta_1 + (\omega_r + 2D) \beta_2 = -\epsilon \beta_2,
\eal
which gives the four eigenenergies
\bal
\epsilon = \pm \sqrt{\omega_r (\omega_r + 4 D \pm 2 J)}.
\eal
Taking into account the constraint conditions in \eq{\ref{eq_cons}}, the four eigenergies are reduced into two elementary excitation energy $\epsilon_{\pm} = \sqrt{\omega_r (\omega_2 + 4 D \pm 2 J)}$ as in \eq{\ref{eq_cons}}. And we obtain the diagonalized Hamiltonian expressed in \eq{\ref{eq_hrdj2}}. Physically, the elementary excitation energy of a system need to be positive real-valued, otherwise the system is unstable. And thus we arrive at the frequency condition that $\omega_2+4D\pm2J \geq 0$ need to be satisfied.

\section*{Appendix B:The effective Hamiltonian}
\label{sec_app2}
The dimensionless Hamiltonian $\m{H}=H/\omega_q$ can be rewritten as
\bal
\label{eq_ahfull}
\m{H} = \sum_{j = 1,2} \left(\m{H}_{0}^j+ \m{H}_{\rm q-r}^j + \m{H}_{\rm A2}^j \right) + \m{H}_{\rm hop},
\eal
where
\bal
\label{eq_ah}
\m{H}_{0}^j &= \frac{a_j^{\dagger}a_j}{\eta}+\frac{1}{2}\sigma_j^z , \\
\m{H}_{\rm q-r}^j &=  \frac{\tilde{g}}{2 \sqrt{\eta}}\sigma_j^x(a_j + a_j^{\dagger}),\\
\m{H}_{\rm A2}^j &=  \frac{D}{\omega_q}  (a_j + a_j^{\dagger})^2, \\
\m{H}_{\rm hop} &= \frac{\tilde{J}}{2\eta}(a_1 + a_1^{\dagger})(a_2 + a_2^{\dagger}).
\eal
For the $j$-th site, the unperturbed Hamiltonian $\m{H}_{0}^j$ has decoupled spin subspaces $\m{H}_{\uparrow}^j$ and $\m{H}_{\downarrow}^j$, respectively. In the limit of $\eta \rightarrow \infty$, the low-lying eigenstates of each $\m{H}_{0}^j$ stay in the spin subspace of $\m{H}_{\downarrow}^j$. However, the interaction between the qubit and the corresponding photonic mode $\m{H}_{\rm q-r}^j$ introduces exchange between the respective two spin subspaces. Therefore, the resulting virtual transitions may revise the properties of the low energy lying eigenstates.

To promote the investigation of the critical features in this model, we derive the low-energy effective Hamiltonian $\m{H}'$ in \eq{\ref{eq_hd_sw}} from the full Hamiltonian $H$ given in \eq{\ref{eq_hd}} of the main text. The idea behind is to find a unitary transformation $U$ which makes the transformed Hamiltonian, $U^\dag \m{H} U$, free of coupling terms between the respective spin subspaces $\m{H}_{\uparrow}^j$ and $\m{H}_{\downarrow}^j$.

We apply a generalized Schrieffer-Wolff transformation $U = \exp{[S]}$, with $S=\sum_{j=1,2}S^j$, to the Hamiltonian \eq{\ref{eq_ahfull}}, where each generator $S^j$ is anti-Hermitian and block-off-diagonal with respect to the corresponding spin subspaces. The transformed Hamiltonian $\m{H}'= U^\dag \m{H} U$ can be expanded as
\bal
\label{eq_ahsw1}
\m{H}'= \sum_{k=0} \frac{1}{k!} {[\m{H},S]}^{(k)},
\eal
where ${[\m{H},S]}^{(k)}\equiv[{[\m{H},S]}^{(k-1)},S]$, with ${[\m{H},S]}^{(0)}\equiv \m{H}$.

Since the field squeezing term $\m{H}_{\rm A2}^j $ and the hopping term $\m{H}_{\rm hop}$ maintain the respective two spin subspaces $\m{H}_{\uparrow}^j$ and $\m{H}_{\downarrow}^j$, the generator is designed such that $[\m{H}_{\rm A2}^j, S]^{(1)}\equiv 0$, and $[\m{H}_{\rm hop}, S]^{(1)}\equiv 0$. Substituting \eq{\ref{eq_ahfull}} into \eq{\ref{eq_ahsw1}}, we then obtain
\bal
\label{eq_ahsw2}
\m{H}'=\sum_{j = 1,2} \left(\sum_{k=0} \frac{1}{k!} {[(\m{H}_0^j+\m{H}_{\rm q-r}^j),S^j]}^{(k)}+ \m{H}_{\rm A2}^j \right) + \m{H}_{\rm hop}.
\eal
In order to eliminate the coupling between the spin subspaces induced $\m{H}_{\rm q-r}^j$, we block-diagonalize the transformed Hamiltonian $\m{H}'$ up to fourth order in $\eta^{-1/2}$ with respect to the spin subspaces. For this purpose, the following conditions need to be satisfied,
\bal
\label{eq_as1}
[\m{H}_0^j,S_1^j] = -\m{H}_{\rm q-r}^j,  \\
[\m{H}_0^j,S_3^j] = -\frac{1}{3}[[\m{H}_{\rm q-r}^j,S_1^j],\;S_1^j],
\eal
where the generator $S^j$ is divided into $S^j= S_1^j + S_3^j$ and the property that $\m{H}_{\rm q-r}^j$ is off-block-diagonal has been employed.
Respecting the above conditions, the generator $S^j$ is found to be
\bal
\label{eq_as2}
S_1^j=-i\frac{\tilde{g}}{2\sqrt{\eta}}(a_j + a^\dag_j)\s_y ,  \\
S_3^j=\frac{i}{6}{\left(\frac{\tilde{g}}{\sqrt{\eta}}\right)}^3 {(a_j + a^\dag_j)}^3\s_y.
\eal
And thus the transformed Hamiltonian is given by
\begin{widetext}
\bal
\label{eq_aheff}
\m{H}'&=\sum_{j = 1,2} \left(\m{H}_0^j+\m{H}_{\rm A2}^j+\frac{1}{2}[\m{H}_{\rm q-r}^j,S_1^j] + \frac{1}{2}[\m{H}_{\rm q-r}^j,S_3^j] - \frac{1}{24}[[[\m{H}_{\rm q-r}^j,S_1^j],S_1^j],S_1^j] \right)  + \m{H}_{\rm hop},\notag \\
&=\sum_{j = 1,2} \left(\frac{a_j^{\dagger}a_j}{\eta}+\frac{1}{2}\sigma_j^z +\frac{\tilde{g}^2(\sigma_j^z+\tilde{D})}{4\eta} (a_j + a_j^{\dagger})^2 - \frac{\tilde{g}^4\sigma_j^z}{16\eta^2} (a_j + a_j^{\dagger})^4\right)+\frac{\tilde{J}}{2\eta}(a_1 + a_1^{\dagger})(a_2 + a_2^{\dagger}),
\eal
\end{widetext}
which is exactly \eq{\ref{eq_hd_sw}} in the main text.




%

\end{document}